\documentclass[aps,prd,onecolumn,superscriptaddress,showpacs,bibnotes]{revtex4}
\usepackage{graphicx}
\usepackage{dcolumn}
\usepackage{bm}
\usepackage{epstopdf}
\usepackage{mathtools}
\usepackage{natbib}
\usepackage{captcont}
\usepackage{xcolor}
\usepackage{ulem}
\usepackage{supertabular}
\usepackage[FIGTOPCAP]{subfigure}

\begin{document}
\title{Late afterglow emission statistics: a clear link between GW170817 and bright short GRBs}
\affiliation{Key Laboratory of Dark Matter and Space Astronomy, Purple Mountain Observatory, Chinese Academy of Sciences, Nanjing 210034, China }
\affiliation{University of Chinese Academy of Sciences, Yuquan Road 19, Beijing, 100049, China}
\affiliation{School of Astronomy and Space Science, University of Science and Technology of China, Hefei, Anhui 230026, China.}
\affiliation{College of Science, Guilin University of Technology, Guilin 541004, China}
\author{Kai-Kai  Duan}
\affiliation{Key Laboratory of Dark Matter and Space Astronomy, Purple Mountain Observatory, Chinese Academy of Sciences, Nanjing 210034, China }
\affiliation{University of Chinese Academy of Sciences, Yuquan Road 19, Beijing, 100049, China}
\author{Zhi-Ping Jin$^\ast$}
\affiliation{Key Laboratory of Dark Matter and Space Astronomy, Purple Mountain Observatory, Chinese Academy of Sciences, Nanjing 210034, China }
\affiliation{School of Astronomy and Space Science, University of Science and Technology of China, Hefei, Anhui 230026, China.}
\author{Fu-Wen Zhang}
\affiliation{College of Science, Guilin University of Technology, Guilin 541004, China}
\author{Yi-Ming Zhu}
\affiliation{Key Laboratory of Dark Matter and Space Astronomy, Purple Mountain Observatory, Chinese Academy of Sciences, Nanjing 210034, China }
\affiliation{School of Astronomy and Space Science, University of Science and Technology of China, Hefei, Anhui 230026, China.}
\author{Xiang Li$^\ast$}
\affiliation{Key Laboratory of Dark Matter and Space Astronomy, Purple Mountain Observatory, Chinese Academy of Sciences, Nanjing 210034, China }
\author{Yi-Zhong Fan$^\ast$}
\affiliation{Key Laboratory of Dark Matter and Space Astronomy, Purple Mountain Observatory, Chinese Academy of Sciences, Nanjing 210034, China }
\affiliation{School of Astronomy and Space Science, University of Science and Technology of China, Hefei, Anhui 230026, China.}
\author{Da-Ming Wei}
\affiliation{Key Laboratory of Dark Matter and Space Astronomy, Purple Mountain Observatory, Chinese Academy of Sciences, Nanjing 210034, China }
\affiliation{School of Astronomy and Space Science, University of Science and Technology of China, Hefei, Anhui 230026, China.}
\begin{abstract}
GW170817, the first neutron star merger event detected by advanced LIGO/Virgo detectors, was associated with an underluminous short duration GRB 170817A. In this work we compare the forward shock afterglow emission of GW170817/GRB 170817A to other luminous short GRBs (sGRBs) with both a known redshift and an afterglow emission lasting at least one day after the burst. In the rapid decay phase, the afterglow emission of the bright sGRBs and GW170817/GRB 170817A form a natural and continuous sequence, though separated by an observation time gap.  If viewed on-axis, the forward shock afterglow emission of GW170817/GRB 170817A would be among the brightest ones detected so far. This provides a strong evidence for the GW170817-like merger origin of bright sGRBs, and suggests that the detection of the forward shock afterglow emission of most neutron star merger events are more challenging than the case of GW170817 since usually the mergers will be more distant and the viewing angles are plausibly higher.
\end{abstract}
\maketitle

The mergers of double neutron star systems or the neutron star-black hole binaries generate strong gravitational wave radiation as well as short duration gamma-ray bursts (sGRBs; including the so-called long-short GRBs) \cite{Eichler1989,Piran2004,Berger2014}. Before 2017, it was widely believed that the GW/sGRB association rate is low since the sGRB outflows are highly collimated with a typical half-opening angle of $\sim 0.1$ rad\cite{Clark2015}. Surprisingly, on 2017 August 17, the gamma-ray monitor onboard the {\it Fermi} $\gamma-$ray space telescope had successfully detected a weak short GRB 170817A\cite{Goldstein2017} that is spatially and temporally correlated with GW170817, the first neutron star merger event detected by advanced LIGO/Virgo\cite{Abbott2017}. The GW/sGRB association has been formally established. However, with a low fluence as well as a short distance ($D\sim 40$ Mpc), the isotropic-equivalent gamma-ray radiation energy of GRB 170817A is just $\sim 3\times 10^{46}$ erg, which is at least 100 times dimmer than that of the typical sGRBs.  An under-luminous sGRB could either result from the breakout of the mildly relativistic shock from the leading edge of the merger-driven quasi-isotropic sub-relativistic ejecta\cite{Kasliwal2017} or be the faint prompt emission of a highly structured relativistic ejecta viewed at a large polar angle\cite{Jin2018}. The puzzling fact that GRB 170817A and the long duration GRB 980425 (at a distance of $D\sim 36$ Mpc and has been suggested to be the shock breakout signal\cite{Kulkarni1998}), the closet two events with remarkably different progenitors, have rather similar luminosity and spectral peak energy\cite{WangH2017}, may favor the shock breakout model. It is thus unclear whether GW170817-like mergers are indeed the sources of the bright sGRBs or not. The forward shock afterglow observations of GW170817/GRB 170817A are helpful in answering such a question. Though the ``early" rising X-ray and radio afterglow emission could be reproduced by a cocoon-like mildly relativistic ejecta\cite{Kasliwal2017}, the late time afterglow data modelings strongly favor the presence of an off-axis relativistic (structured) outflow component \cite{Yue2018,D'Avanzo2018,Lamb2018,Mooley2018b}. Particularly, the off-axis relativistic outflow component at a viewing angle of $\theta_{\rm v} \sim 0.35$ rad has been convincingly identified/measured in the radio image \cite{Mooley2018}. Nevertheless, a direct ``observational" link between GW170817/GRB 170817A and bright sGRBs is still lack. In this work, we carry out statistical studies of the sGRB afterglow data and aim to establish such a connection.

In the fireball afterglow model, the emission arises from the shock-accelerated electrons (with an energy distribution power law index $p$) moving in the shock-generated magnetic fields \cite{Piran2004}. A simplified uniform energy distribution with an abrupt energy
depletion of a conical ejecta has been assumed in most studies.
In reality, the energy distribution function could be more complicated and several empirical structured jet models have been proposed in the literature  \cite{Dai2001,Rossi2002,Berger2003,Zhang2004,Jin2007}. Usually the energy distribution is insensitive on the polar angle for $\theta\leq \theta_{\rm c}$ but drops rapidly outward, where $\theta_{\rm c}$ is the half opening angle of the energetic core. The GRBs viewed at the angles of  $\theta_{\rm v}\leq \theta_{\rm c}$ are called as the on-axis events and otherwise the off-axis events.
The afterglow emission of structured jets have been extensively calculated in the literature. If viewed off-axis, it is found that the early afterglow emission are sensitively dependent on the viewing angle $\theta_{\rm v}$ (the larger $\theta_{\rm v}$, the much weaker the emission), while at late times with the considerably decreased bulk Lorentz factor the viewing angle effect will be significantly suppressed \cite{Wei2003,Kumar2003,Lamb2017}. In particular, a quick decline ($t^{-p}$) phase will appear in the afterglow lightcurve when the bulk Lorentz factor of the ejecta drops to $\sim 1/(\theta_{\rm v}+\theta_{\rm c})^{-1}$, after which the afterglow emission viewed at different $\theta_{\rm v}$ will be similar\cite{Wei2003,Kumar2003,Lamb2017}. This conclusion holds for the off-axis uniform ejecta as well. Therefore we can ``extrapolate" the very-late quick-decaying X-ray and optical afterglow emission of GW170817/GRB 170817A to $t_{\rm com}\sim 2$ days (i.e., if viewed on-axis) after the burst and then compare them to other distant sGRBs (Please see Fong et al.\cite{Fong2017} instead for a direct comparison of the ``early" emerging forward shock emission of GW170817/GRB 170817A to the afterglow emission of other sGRBs). The choice of such a $t_{\rm com}$ is for two reasons. One is that at such a late time the forward shock emission is usually in the post-jet-break phase if viewed on-axis (i.e., the bulk Lorentz factor of the decelerated ejecta $\Gamma$ drops below $1/\theta_{\rm c}$ for $\theta_{\rm c}\sim 0.1$. Note that the energetic core of the relativistic outflow driven by GW170817 has an $\theta_{\rm c} \approx 0.08$ rad \cite{Mooley2018}). The other is that the afterglow lightcurves of some sGRBs do not cover a longer time. For the radio emission, $t_{\rm com}\sim 10$ days is needed otherwise the typical synchrotron radiation frequency of the forward shock electrons ($\nu_{\rm m}$, which is independent of the number density of circum-burst medium) is still above the observer's frequency and the flux will not drop with time quickly \cite{Piran2004}. If the burst was born in a dense circum-burst medium, the synchrotron self-absorption plays an important role in suppressing the radio emission, too. Fortunately, for the sGRBs, usually the medium density is low and the self-absorption correction is ignorable.

For our purpose,
we select the events with both a known redshift and an afterglow emission lasting at least one day after the burst. In comparison to the long-duration GRBs, sGRBs have smaller $E_{\rm k}$ mainly due to the shorter durations. The number density of the medium surrounding the sGRBs is also expected to be lower. That is why usually the sGRB forward shock afterglow emission is faint and can not be detected in a long term \cite{Fong2015}. Most X-ray data were recorded by the {\it Swift} X-ray Telescope (XRT) and are available at http://www.swift.ac.uk/xrt$_{-}$curves/ and http://www.swift.ac.uk/xrt$_{-}$spectra/ \cite{Evans2009}. For GRB 050709, GRB 050724, GRB 051221A, GRB 060505, GRB 120804A, GRB 130603B, GRB 140903A and GRB 150101B, the X-ray data from Chandra or XMM-Newton satellites are available\cite{Fong2015}. The optical observations were performed by various telescopes, while at late times only a few very-large ($\sim 8-10$m) ground-based telescopes and the Hubble Space Telescope (HST) are able to contribute. Our X-ray sample consists of $19$ bursts. While in optical and radio bands, there are just $7$ and $3$ bursts in our samples, respectively. The details of our samples and the data sources (references) are introduced in the Appendix.

\begin{figure}[!ht]
\centering
\includegraphics[width=0.8\textwidth]{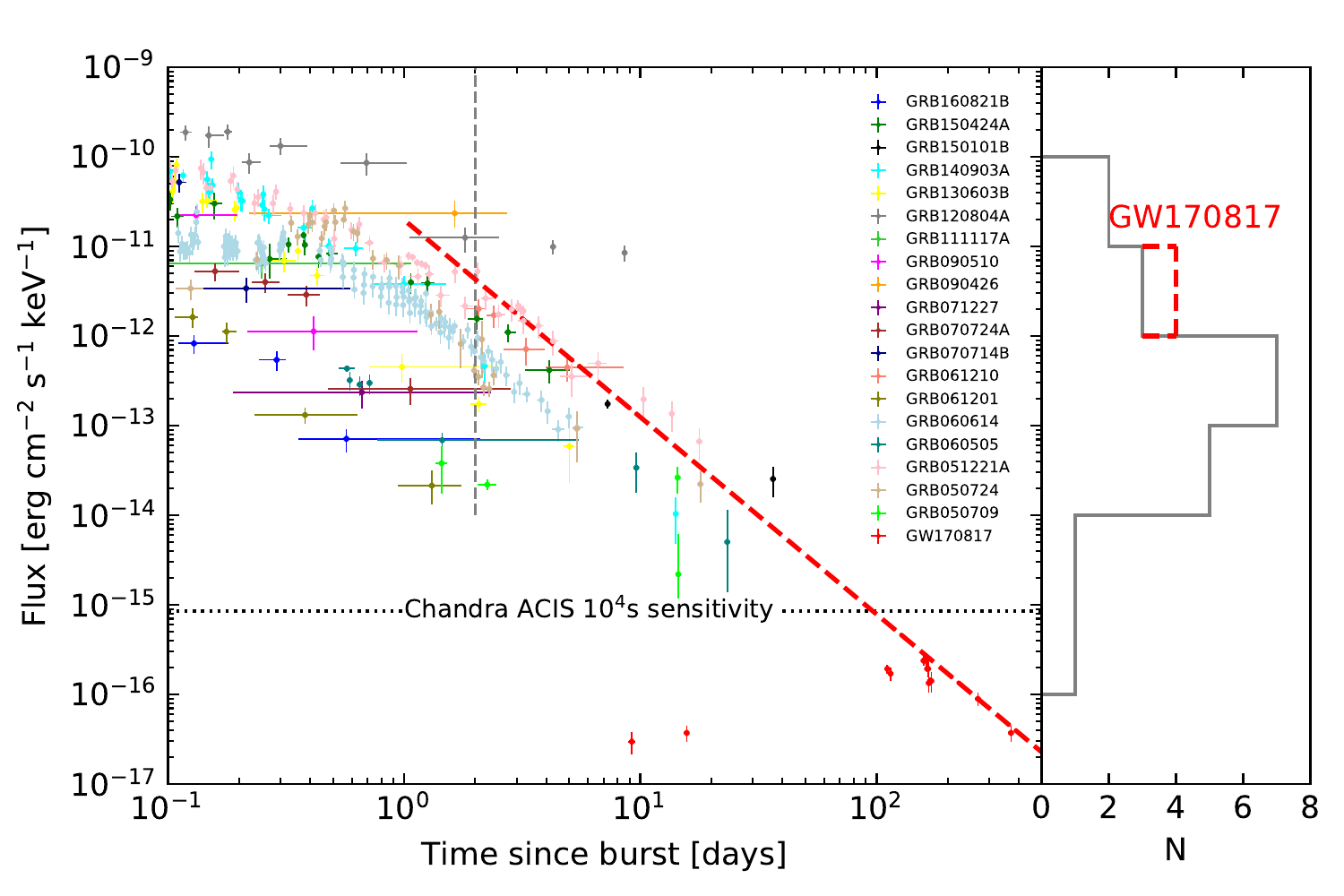}
\caption{{ The ``long-lasting" X-ray (1.732 keV) afterglow emission of some sGRBs and GW170817/GRB 170817A\cite{D'Avanzo2018,Troja2018}, if occurred at the same distance of $200$ Mpc.} The red dashed line represents the ``on-axis" extrapolation to early times from the very late ($t>150$ day) X-ray afterglow data of GRB 170817A, the vertical dashed line represents the time of 2 day after the burst, and the dotted horizontal line is the Chandra ACIS $10^{4}$ s observation sensitivity (http://cxc.harvard.edu/cdo/about$_{-}$chandra/). Please see the Appendix  for the details of the X-ray sample. The right panel presents the distribution of the X-ray fluxes at a fixed time of 2 days after the burst.
}
\label{fig:X-ray}
\end{figure}

\begin{figure}[!ht]
\centering
\includegraphics[width=0.8\textwidth]{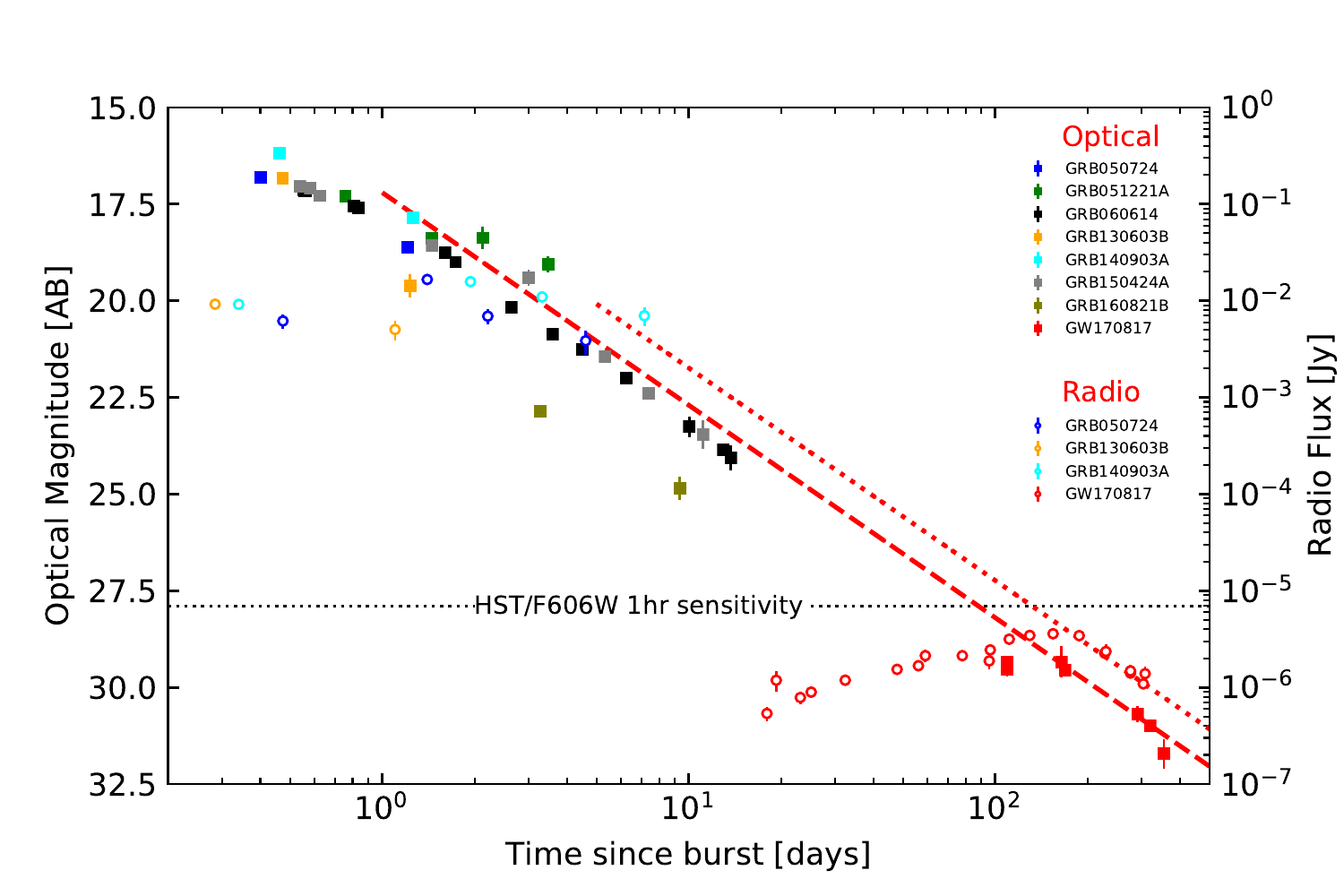}
\caption{{ The ``long-lasting" $R-$band (the filled squares) and radio (6 GHz; the open circles) afterglow emission of some sGRBs and GW170817/GRB 170817A, if took place at the same distance of $200$ Mpc.}  The HST sensitivity is from the Wide Field Camera 3 Instrument Handbook for Cycle 27 (http://www.stsci.edu/hst/wfc3/design/at\_a\_glance/documents/handbooks/currentIHB/wfc3\_ihb.pdf). The forward shock optical and radio afterglow data of GW170817/GRB 170817A are adopted from the literature\cite{Lamb2018,Lyman2018,Mooley2018,Mooley2018b}. The details of the optical and radio samples are presented in the Appendix.
}
\label{fig:optical}
\end{figure}

\begin{figure}[!ht]
\includegraphics[width=0.8\textwidth]{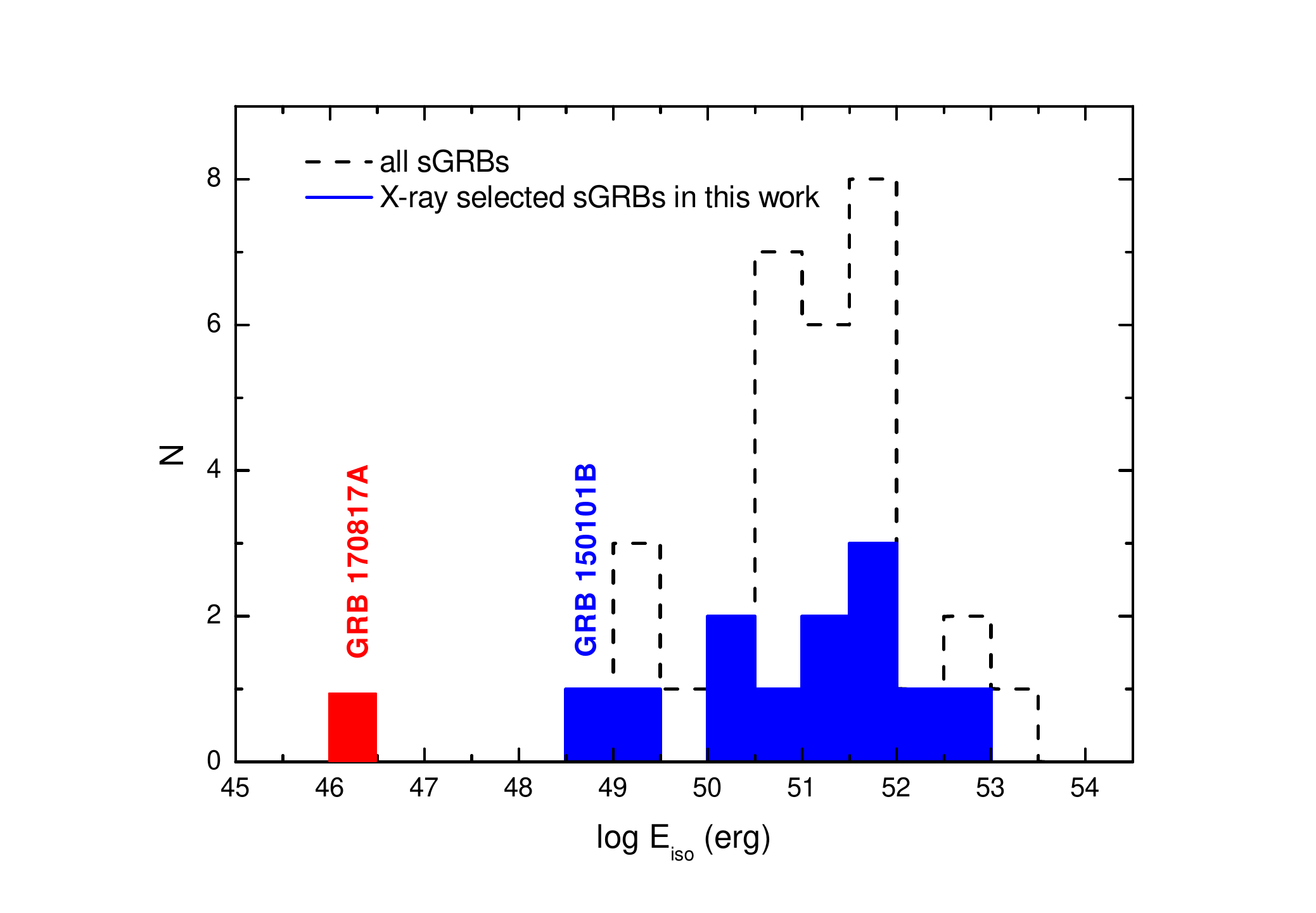}
\caption{{ The distribution of the isotropic gamma-ray energy $E_{\rm iso}$ of sGRBs with a known redshift and a reliably measured spectrum.} The black dashed histogram represents all current short bursts, while the red and purple shaded histograms represent the events with ``long-lasting" X-ray and optical afterglow emission, respectively. GRB 170817A is marked in blue. The values of $E_{\rm iso}$ are calculated in the rest frame energy band of $1-10^{4}$ keV, where the redshift and spectral information of the bursts are adopted from the literature \cite{Berger2014,Fong2015,Lien2016,NarayanaBhat2016,Tsvetkova2017,Goldstein2017,Troja2018}.  }
\label{fig:eiso}
\end{figure}

In Fig.\ref{fig:X-ray} and Fig.\ref{fig:optical}  we show the X-ray (1.732 keV), optical ($R-$band) and radio (6 GHz) fluxes if observed at a distance of $200$ Mpc, motivated by the fact that the averaged sensitive range of the advanced LIGO/Virgo detectors in their full-sensitivity run is about 210 Mpc, for the current samples, respectively. Due to the faintness of the sGRB afterglow emission, there are gaps of the data between the previous more-distant events and GW170817/GRB 170817A (please note that for the latter we only consider the quick decline phase since the early part is significantly influenced by the beam effect of the off-axis outflow). Therefore we extrapolate the very late ($t>200$ day) X-ray and optical afterglow data of GRB 170817A to $t\sim 2$ day after the burst and then compare them to other events. The radio to X-ray spectrum of the forward shock afterglow emission of GW170817/GRB 170817A is $f_\nu \propto \nu^{-0.6}$, which yields a $p=2.2$ in the slow-cooling synchrotron radiation scenario \cite{Lamb2018,Troja2018}. In the jet model, such a $p$ can also reasonably account for the very late flux decline of\cite{Lamb2018} $f\propto t^{-2.42\pm 0.2}$. The extrapolation function of the forward shock emission of GRB 170817A to early times is thus taken as $f\propto t^{-2.2}$. Surprisingly, the forward shock afterglow emission of GW170817/GRB 170817A, the first neutron star merger event detected by advanced LIGO/Virgo, are among the brightest ones for all short GRBs detected so far. Just a few events have X-ray afterglow emission brighter than
that of GRB 170817A, as demonstrated in the right panels of Fig.\ref{fig:X-ray}.
The same conclusion holds for the optical and radio afterglow data as well, though these two samples are rather limited.
We have also compared the distribution of the isotropic gamma-ray energy $E_{\rm iso}$, calculated in the rest frame energy band of $1-10^{4}$ keV, for the sGRBs with well measured spectra
and found no significant difference for the sGRBs with and without ``long-lasting" afterglow emission (see Fig.\ref{fig:eiso}; where the number of events for the X-ray sample are smaller than that presented in Fig.\ref{fig:X-ray} because some bursts are lack of reliable spectral measurements). GRB170817A and GRB 150101B, two short events with the weakest detected prompt emission, have ``bright" late time afterglow emission because of their off-axis nature.

The above results have two intriguing implications. One is that there is a tight connection between GW170817-like mergers and the bright sGRBs though the physical process giving rise to GRB 170817A is still to be pined down. The other is that the detection of the forward shock afterglow emission of most (though not all) neutron star merger events is likely more challenging than in the case of GW170817 since usually the mergers will be more distant and the viewing angles will be larger \cite{Lamb2017}.

In both Fig.\ref{fig:X-ray} and Fig.\ref{fig:optical}, there are observation time gaps (roughly from $\sim 10-30$ days to $\sim 150$ days) between the sharp decline phases of the forward shock afterglow emission of GW170817/GRB 170817A and those of the much more distant events (note that the second nearest short/long-short burst has a redshift about ten times that of GRB170817A). Such gaps are expected to be bridged as some other off-axis GW/GRB events, but less extreme than GW170817/GRB 170817A (i.e., $\theta_v\gtrsim \theta_{\rm c}$, which we call the quasi on-axis events), have been discovered and densely followed. The quasi on-axis events are less frequent than GW170817/GRB 170817A plausibly by a factor of $\lesssim 10$, but statistically the forward shock peak time will be earlier and the afterglow emission are brighter, which can be well recorded. This would be particularly the case in X-ray and radio bands, for which the contamination by the macronova/kilonova are negligible.

With a local neutron star merger rate of $\sim 10^{3}~{\rm Gpc^{-3}}~{\rm yr^{-1}}$, as inferred from both the gravitational wave data \cite{Abbott2017} and the sGRB observations \cite{Fong2015,Jin2018}, in the era of the full-sensitivity run of the second generation gravitational wave detectors, a sample consisting of $\sim 10^{2}-10^{3}$ neutron star mergers will be available. Some of them may generate detectable very-late forward shock afterglow emission. With a reasonably large sample, the luminosity function of the quickly decaying (i.e., post jet-break) forward shock afterglow emission driven by the neutron star mergers will be reconstructed. Intriguing difference between the double neutron star merger events and the neutron star-black hole merger events may be identified in the afterglow data. As for the double neutron star merger events, special attention may be paid on probing the possible correlations between the total mass ($M_{\rm tot}$) or the mass asymmetry ($q$) of the progenitor stars and the luminosity of the very-late afterglow emission. GW170817 has a $M_{\rm tot}\approx 2.74M_\odot$ ($q\approx 0.86$), which seems to be in the high total mass (high mass asymmetry) part of the double neutron star binary systems detected in the Galaxy, while the absence of $M_{\rm tot}$ and $q$ of the progenitor stars for all other sGRBs hamper us to go further. The situation will change dramatically in the next decade and the roles of properties of the progenitor stars on launching the relativistic outflows and generating the (very late) forward shock afterglow emission will be revealed.


{\it Acknowledgments.}
This work was supported in part by NSFC under grants of No. 11525313 (i.e., Funds for Distinguished Young Scholars), No. 11433009, No. 11773078 and 11763003, the Foundation for Distinguished Young Scholars of Jiangsu Province (No. BK20180050), the Chinese Academy of Sciences via the Strategic Priority Research Program (Grant No. XDB09000000) and and the International Partnership Program of Chinese Academy of Sciences (114332KYSB20170008). F.-W.Z. also acknowledges the support by the Guangxi Natural Science Foundation (No. 2017GXNSFAA198094)\\

$^\ast$Email: jin@pmo.ac.cn (Z.P.J), xiangli@pmo.ac.cn (X.L), yzfan@pmo.ac.cn (Y.Z.F).




\clearpage
\begin{center}
{\Large Appendix} \\
\end{center}

\begin{appendix}


Here we introduce our samples and the data sources (references). Note that GW170817/GRB 170817A is excluded in these samples.

{\bf The X-ray sample.} Our X-ray sample consists of 19  events. Most data were recorded by the {\it Swift} X-ray Telescope (XRT) and are available at http://www.swift.ac.uk/xrt$_{-}$curves/ and http://www.swift.ac.uk/xrt$_{-}$spectra/ \cite{Evans2009}. For some bursts of interest, there were deep Chandra or XMM-Newton detections. These bursts include GRB 050709 \cite{Fox2005}, GRB 050724 \cite{Berger2005,Grupe2006}, GRB 051221A \cite{Soderberg2006}, GRB 060505 \cite{Ofek2007}, GRB 120804A \cite{Berger2013a}, GRB 130603B \cite{Fong2014}, GRB 140903A \cite{Troja2016} and GRB 150101B \cite{Fong2015}.

{\bf The optical sample.} The optical sample is composed of 7 events. This sample is significantly smaller than the X-ray sample, likely due to the lack of deep follow-up observations in many events and/or the presence of serious dust extinction. For instance, at $t>0.1$ day after the burst the X-ray afterglow emission of GRB 120804A is brightest in the sample (see Fig.\ref{fig:X-ray}) but no optical emission was detected \cite{Berger2013a}. For GRB 130603B and GRB 140903A, the dust extinctions of the host Galaxies are serious in the optical bands and have been corrected \cite{Fong2014,deUgartePostigo2015,Troja2016}. Other events are GRB 050724 \cite{Berger2005,Malesani2007}, GRB 051221A \cite{Soderberg2006}, GRB 060614 \cite{Della2006,Gal-Yam2006,Yang2015}, GRB 150424A \cite{Jin2018,Knust2017} and GRB 160821B \cite{Jin2018}. For a few bursts such as GRB 050709 and GRB 150101B, optical emission were detected at $t\geq 1$ day. These events, however, are excluded in the current sample since their late time optical emission are likely dominated by the macronova/kilonova component \cite{Jin2016,Troja2018}.
For the same concern, the very late time I/F814W or F160W data of GRB 060614 (for this burst the macronova identification strongly favors a neutron star merger origin though its duration is apparently long \cite{Yang2015}) and GRB 130603B are excluded. In Fig.\ref{fig:optical}, the Galactic extinction corrections have been made for all bursts\cite{Schlafly2011}.

{\bf The radio sample.} So far there are six sGRBs with detected radio emission \cite{Fong2017}.  The radio emission of GRB 051221A was detected but it is likely from the reverse shock rather than the forward shock \cite{Soderberg2006}. GRB 150424A and GRB 160821B had been detected in radio at early times \cite{Fong2015-radio,Fong2016-radio}, but they have not been formally published, yet. Therefore, our radio sample just consists of GRB 050724 \cite{Berger2005,Malesani2007}, GRB 130603B \cite{Fong2014}, and GRB 140903A \cite{Troja2016}.



\end{appendix}

\end{document}